# Early AI Literacy in Culturally Responsive STEM Outreach for Black Youth

*Qusay H. Mahmoud, Kimberly Davis, Paula Duru, Laura Thursby, Hossam Kishawy*
Faculty of Engineering and Applied Science
Ontario Tech University, Oshawa, Ontario L1G 0C5 Canada
{qusay.mahmoud,kimberly.davis,paula.duru,laura.thursby,hossam.kishawy}@ontariotechu.net

*This paper was accepted to the 2026 Canadian Engineering Education Association Conference, but was later withdrawn prior to publication.*

**Abstract** – *Persistent inequities in STEM education continue to limit the participation of Black youth in science and technology fields across Canada. Structural barriers, underrepresentation, and limited access to culturally affirming learning spaces can restrict both opportunity and confidence in pursuing STEM pathways. This paper examines Ontario Tech University's Engineering Outreach Black Youth Program as an exploratory, practice-based case study of culturally responsive STEM outreach. The program creates inclusive environments where Black youth engage in hands-on, culturally grounded STEM experiences supported by mentorship, representation, and community connection. Its recent integration of artificial intelligence (AI) literacy reflects a growing recognition that early engagement with emerging technologies may expand access to future STEM learning opportunities. The paper discusses how AI-focused activities were introduced within this outreach model and examines short-term outcomes related to AI knowledge, confidence, and critical awareness. Findings suggest gains across these areas, while highlighting the need for future research to examine longer-term outcomes related to STEM belonging, identity, and persistence.*

## 1. INTRODUCTION

As artificial intelligence becomes increasingly embedded in education, work, and everyday decision-making, early AI literacy is becoming an important dimension of equitable STEM participation. For Black youth, access to AI learning must involve more than exposure to technical tools; it must also include opportunities to question, evaluate, and shape technologies that may affect their communities. Recent AI literacy frameworks emphasize that young people should not only use AI tools, but also understand, evaluate, and critically examine their social and ethical implications [12], [13].

Within this context, Ontario Tech University's Engineering Outreach Black Youth Program provides a useful setting for examining how culturally responsive STEM outreach can introduce AI literacy in ways that are accessible, relevant, and identity affirming. The program builds on mentorship, representation, hands-on learning, and community connection to support Black youth engagement in STEM. Its integration of AI-focused activities extends this outreach model into an emerging area of technological learning where questions of access, bias, representation, and agency are especially important. The goal is to embed early AI literacy within a culturally responsive outreach model that supports Black youth in engaging critically and confidently with new technologies. The guiding inquiry question is: How can culturally grounded STEM outreach introduce AI literacy in ways that support engagement, confidence, and early technological agency among Black youth?

This paper presents an exploratory, practice-based case study examining how culturally responsive STEM outreach may support early AI literacy and confidence among Black youth in informal learning environments. Using the Engineering Outreach Black Youth Program as a case study, the paper explores how critically informed, age-appropriate AI learning experiences may support engagement, confidence, and early indicators of technological agency among participants. To support this discussion, the paper examines strategies and challenges of integrating AI-focused learning into outreach contexts, outlines methods for making complex AI concepts accessible to K–12 learners, and discusses participant learning outcomes and engagement feedback.

The Black Youth Program offers workshops, mentorship, and hands-on activities designed and led by Black professionals and post-secondary students. Positioning Black mentors as both technical guides and visible role models, the program intentionally centers representation, relational learning, and community connection as core pedagogical strategies. Through sustained hands-on engagement and mentorship, the program aims to support the development of technical and problem-solving skills alongside confidence, curiosity, and positive experiences within STEM learning environments.

As technology continues to evolve, the program remains dedicated to staying current with emerging trends in education and innovation, particularly the growing influence of artificial intelligence across STEM fields and everyday life. This commitment is reflected in the introduction of AI-focused content, including foundational lessons on what artificial intelligence is, how it appears in everyday life, and basic concepts of machine learning informed by emerging K–12 AI competency frameworks [17]. These concepts are taught through interactive, age-appropriate activities, such as exploring facial recognition, understanding algorithmic bias, and discussing how AI systems make decisions.

The approach draws on Culturally Relevant Pedagogy, which emphasizes identity affirmation in learning contexts [1], and on emerging frameworks in AI and critical technological literacy that foreground both technical understanding and critical engagement with AI systems [2], [3]. By situating AI education within culturally affirming contexts, the program contributes to scholarship suggesting that early, inclusive interventions may support more equitable participation in AI and STEM pathways [4], [5]. While existing STEM outreach research emphasizes the importance of culturally affirming learning environments, fewer practice-based studies have examined how these approaches can be adapted specifically for early AI literacy education. This gap is particularly important given growing concerns that inequities in STEM participation may also extend into emerging technological domains such as artificial intelligence.

To this end, the paper is organized as follows. Section 2 reviews related literature on STEM inequities, culturally responsive outreach, AI literacy, and virtual STEM learning. Section 3 describes the outreach framework, workshop implementations, and evaluation methods. Section 4 presents findings on AI literacy, confidence, engagement, and ethical understanding. Section 5 discusses implications, limitations, and future directions.

# 2. LITERATURE REVIEW

Existing research across STEM education, culturally responsive outreach, and emerging work on AI literacy provides important context for this study. Together, these bodies of scholarship highlight both the persistence of inequities in STEM participation among Black youth and the need for early, inclusive approaches to technological learning as artificial intelligence becomes increasingly embedded across STEM fields.

## 2.1. Inequities in STEM Participation

Research consistently documents longstanding inequalities in access, representation and outcomes for Black students across STEM education pathways. Studies have identified barriers, including underrepresentation and a lack of diverse role models [4], [6], [7], limited access to advanced enrichment opportunities [8], deficit-oriented narratives, and the absence of culturally affirming narratives that reflect the lived experiences of Black youth [9], [10], [11]. These factors contribute not only to disparities in academic outcomes but also to diminished STEM identity, confidence, and a sense of belonging among Black youth. Although several studies in this area focus specifically on Black girls in STEM, these findings remain relevant to broader discussions of representation, belonging, and culturally responsive outreach among Black youth more generally.

## 2.2. Culturally Responsive STEM Outreach

In response to these inequities, a growing body of research highlights the role of culturally responsible, community-based STEM outreach in supporting engagement and persistence among underrepresented youth. Studies suggest that when STEM learning is situated within culturally affirming contexts, students are more likely to develop a sense of belonging and sustained interest in STEM education [1], [5], [10]. Related work on STEM counterspaces further suggests that culturally affirming learning environments can support students from historically marginalized groups by fostering belonging, challenging deficit narratives, and creating spaces where students' identities and experiences are recognized as assets [14]. These studies highlight the importance of early experiences in shaping students' perceptions of who belongs in STEM and who can succeed in technical fields. Early exposure to relatable role models, hands-on learning, and inclusive environments provides opportunities to see oneself reflected in STEM spaces.

## 2.3. AI Literacy

As artificial intelligence increasingly shapes STEM disciplines and everyday life, these same mechanisms (identity affirmation, representation, and early engagement) are central to how young people encounter and engage with AI. Without intentional intervention, longstanding inequities in STEM participation risk being reproduced within emerging technological domains. Within this context, emerging scholarship has begun to consider how early AI literacy, particularly when approached through culturally responsive frameworks, can support more equitable engagement with AI for Black youth. For instance, Atias and Mawasi provide a systematic review of AI literacy programs for children and youth, finding that while many initiatives emphasize technical and operational competencies, sociocultural and critical dimensions of AI literacy remain comparatively underdeveloped [3]. Recent AI literacy frameworks further emphasize that AI education should include not only technical understanding, but also critical evaluation, responsible use, human judgment, and attention to justice, bias, and societal impact [12], [13]. This broader framing is especially important for Black youth, as AI systems can reproduce inequities when issues such as dataset representation, algorithmic bias, and social impact are not explicitly addressed. Further, Lee et al. show that while AI and emerging technologies hold promise for supporting inclusion, their impact on minoritized students depends heavily on sociocultural context and the extent to which educational approaches to AI attend to students' lived experiences and identities [2]. These findings are particularly salient for equity-focused outreach, as they suggest that current AI literacy efforts may insufficiently address identity, power, and lived experience in ways that support meaningful engagement for Black youth.

### 2.4. Virtual and Informal STEM Learning

This scholarship remains at an early stage, particularly with respect to practice-based research examining how culturally grounded outreach initiatives can integrate AI learning to support engagement, belonging, and technological agency among Black youth. This paper contributes to this emerging area of scholarship by examining the Engineering Outreach Black Youth Program as a case study, offering practice-informed insights into how early AI literacy can be meaningfully introduced within culturally affirming STEM outreach contexts. Because one implementation examined in this study occurred through a virtual AI Innovation Club, research on online STEM outreach is also relevant to the intervention design. Recent studies on virtual STEM camps suggest that online outreach can support engagement and participation when intentionally structured around near-peer mentorship, collaborative interaction, guest speakers, and active learning opportunities [15]. Virtual delivery alone, however, does not automatically foster belonging, highlighting the importance of culturally responsive facilitation and relational engagement in online STEM learning environments.

## 3. APPROACH

This study was guided by a culturally responsive AI education framework designed to make artificial intelligence learning accessible, relevant, and culturally affirming for youth in informal STEM outreach settings. The framework was implemented through adaptive, hands-on AI workshops co-designed by university outreach educators, community partners, and mentors to ensure alignment with participants' lived experiences, cultural contexts, and prior exposure to STEM. Rather than relying on a fixed curriculum, workshop content was intentionally flexible and iteratively refined to maintain relevance for diverse cohorts and returning participants.

Workshops embedded representation, real-world relevance, and identity-informed learning into AI activities to support technical engagement and create conditions associated with belonging in STEM. For instance, during an AI-focused STEM Challenge, students researched the contributions of Black innovators, used Google Teachable Machine to reimagine historic inventions through machine learning applications, and pitched AI-driven app concepts demonstrating contemporary relevance. This approach positioned participants as creators and innovators while explicitly connecting AI learning to cultural identity, creativity, and problem-solving.

The workshop design also drew on critical AI literacy frameworks that emphasize understanding, evaluating, and using AI while critically examining bias, representation, human judgment, and societal impact [12], [13]. Unlike introductory AI outreach models that focus primarily on technical skill acquisition, the program intentionally integrated mentorship, identity affirmation, cultural relevance, and discussions of ethics and algorithmic bias alongside technical learning activities. This approach also aligned with counterspace-oriented perspectives in STEM outreach by positioning Black mentors, cultural identity, and community relevance as central features of the learning environment [5], [14]. For the virtual AI Innovation Club, the instructional structure incorporated practices associated with effective online STEM outreach, including near-peer mentorship, collaborative discussion, and interactive learning activities [15].

### 3.1. AI Learning Activities and Implementation

Across the broader Black Youth program, participants engaged in experiential AI activities demonstrating real-world applications of artificial intelligence and aligned with emerging K–12 AI competency development approaches [17]. These included training machine learning models using Google Teachable Machine, ideation with generative AI tools (e.g., ChatGPT), robotics programming with mBots for line-following and obstacle avoidance, AI-based creative platforms such as Quick, Draw!, and media literacy exercises focused on distinguishing AI-generated from human-created content. Collectively, these activities emphasized conceptual understanding of how AI systems function alongside critical awareness of their limitations, biases, and ethical implications, including responsible and safe AI use. Workshops were designed to be collaborative, age-appropriate, and inquiry-driven, prioritizing active learning and technological self-efficacy.

Implementation followed an iterative design process informed by facilitator reflections, participant engagement patterns, and informal community feedback. Mentorship was embedded as a central pedagogical component to support sustained engagement, identity affirmation, and increased confidence in STEM and AI learning environments. This iterative, mentorship-driven structure is a key feature of the culturally responsive framework, enabling instructional strategies to evolve in response to learner needs and contextual factors.

### 3.2. Evaluation Methods

The study in this paper is exploratory in nature and designed primarily to examine short-term learning outcomes and participant experiences within informal outreach settings. Survey measures focused primarily on AI literacy and confidence outcomes and did not directly assess longer-term constructs such as STEM identity formation, sustained belonging, or persistence intentions. To evaluate these practices, a brief pre–post survey was administered in two selected implementations within the broader program: (1) a virtual Black Youth AI Innovation Club consisting of three 2-hour sessions (n = 14), and (2) a 45-minute in-person Black Youth AI workshop delivered to a separate cohort (n = 10). These implementations were chosen to examine learning outcomes across different instructional durations while maintaining consistency in survey instrumentation.

The survey was intentionally designed to be short, youth-friendly, and suitable for informal learning environments to maximize completion rates. Items measured AI knowledge, ability to explain AI concepts, confidence in using AI, perceived usefulness of AI, and ethical awareness (e.g., understanding that AI can make mistakes). Surveys were administered digitally at the beginning and end of each program, and responses were anonymously linked using participant-generated codes to enable paired analysis while preserving confidentiality and avoiding the collection of personally identifiable information. Only matched pre–post responses were included in the analysis.

Quantitative analysis focused on within-subject pre–post comparisons using descriptive statistics and normalized learning-gain calculations across constructs. Because survey items were measured on different Likert scales, responses were normalized to a standard 0–100 scale to ensure comparability. Given the exploratory nature of the outreach context and the small sample size, the analysis emphasizes descriptive trends in AI literacy, confidence, and conceptual understanding rather than making generalized conclusions. In addition to quantitative responses, optional open-ended survey comments and participant feedback were collected and used as illustrative supporting evidence to contextualize learning experiences, rather than as a primary source of formal qualitative analysis.

By integrating culturally responsive pedagogy, adaptive curriculum design, and lightweight evaluation methods, this approach enabled assessment of measurable AI literacy development, confidence, and participant reflections on inclusion and engagement within AI and STEM learning environments.

## 4. RESULTS AND DISCUSSION

### 4.1. Overview

A total of 37 youth participated in the AI programs (25 in the virtual AI Innovation Club and 12 in the in-person AI workshop). However, the present analysis is based on matched pre–post survey responses (n = 24), including 14 participants from the virtual club and 10 from the 45-minute in-person workshop, to ensure accurate within-subject comparison of learning outcomes (Table 1).

**Table 1:** Participants overview.

| Measure | Innovation Club | Workshop |
|---|---|---|
| Total Participants | 25 | 12 |
| Matched Responses | 14 | 10 |
| Grade Range | 6 - 12 | 8 - 12 |
| Prior AI Use (%) | 92.86% | 100% |

Identical pre–post surveys were administered across both implementations and normalized to a 0–100 scale to ensure comparability across Likert items measured on different scales. The surveys assessed AI knowledge, the ability to explain AI concepts, and confidence in using AI, along with post-survey items on conceptual and ethical understanding.

### 4.2. Learning Gains Across Implementations

Across both implementations, participants showed gains in AI knowledge, conceptual understanding, and confidence using AI tools (Figure 1). These gains suggest that the culturally responsive AI framework supported short-term learning across both virtual and in-person contexts.

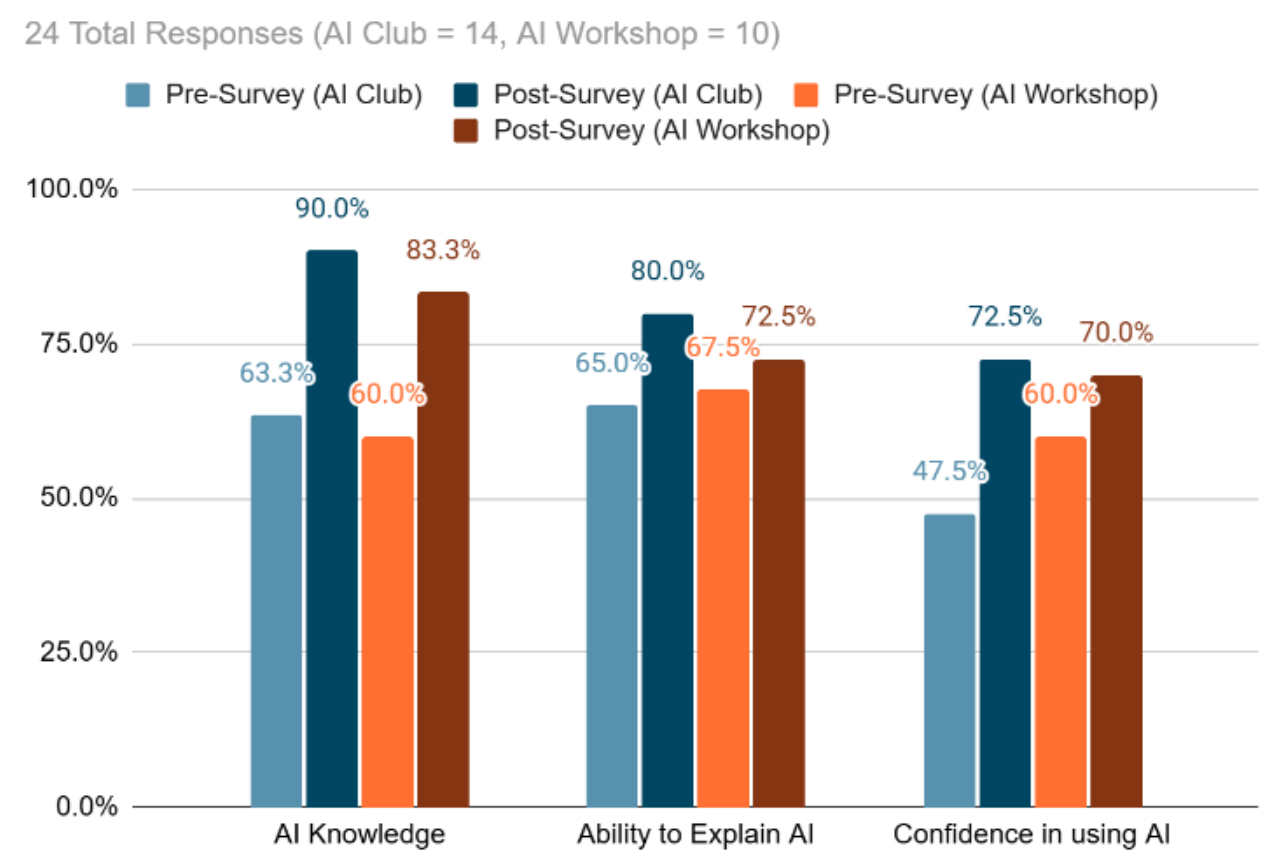

**Figure 1.** AI literacy pre- and post-survey outcomes across two implementations.

Participants in the virtual AI Innovation Club showed substantial gains across all core constructs: AI knowledge increased from 63.3% to 90.0% (+26.7%), ability to explain AI concepts from 65.0% to 80.0% (+15.0%), and confidence from 47.5% to 72.5% (+25.0%). These results suggest that sustained, hands-on engagement with culturally relevant AI activities supported conceptual learning and technological self-efficacy, as reflected in one participant's comment: "It was cool learning how AI works, what it can do, and experimenting with AI!"

Participants in the 45-minute AI workshop also showed gains, though more modest: AI knowledge increased from 60.0% to 83.3% (+23.3%), ability to explain AI from 67.5% to 72.5% (+5.0%), and confidence from 60.0% to 70.0% (+10.0%). The larger gains in the multi-session format suggest that extended exposure may support deeper learning, while the shorter workshop results indicate potential scalability in time-constrained outreach settings.

### 4.3. Shifts in Knowledge and Confidence

Distributional results (**Error! Reference source not found. & Error! Reference source not found.**) corroborate the normalized learning gains by showing clear shifts toward higher knowledge and confidence categories across both implementations.

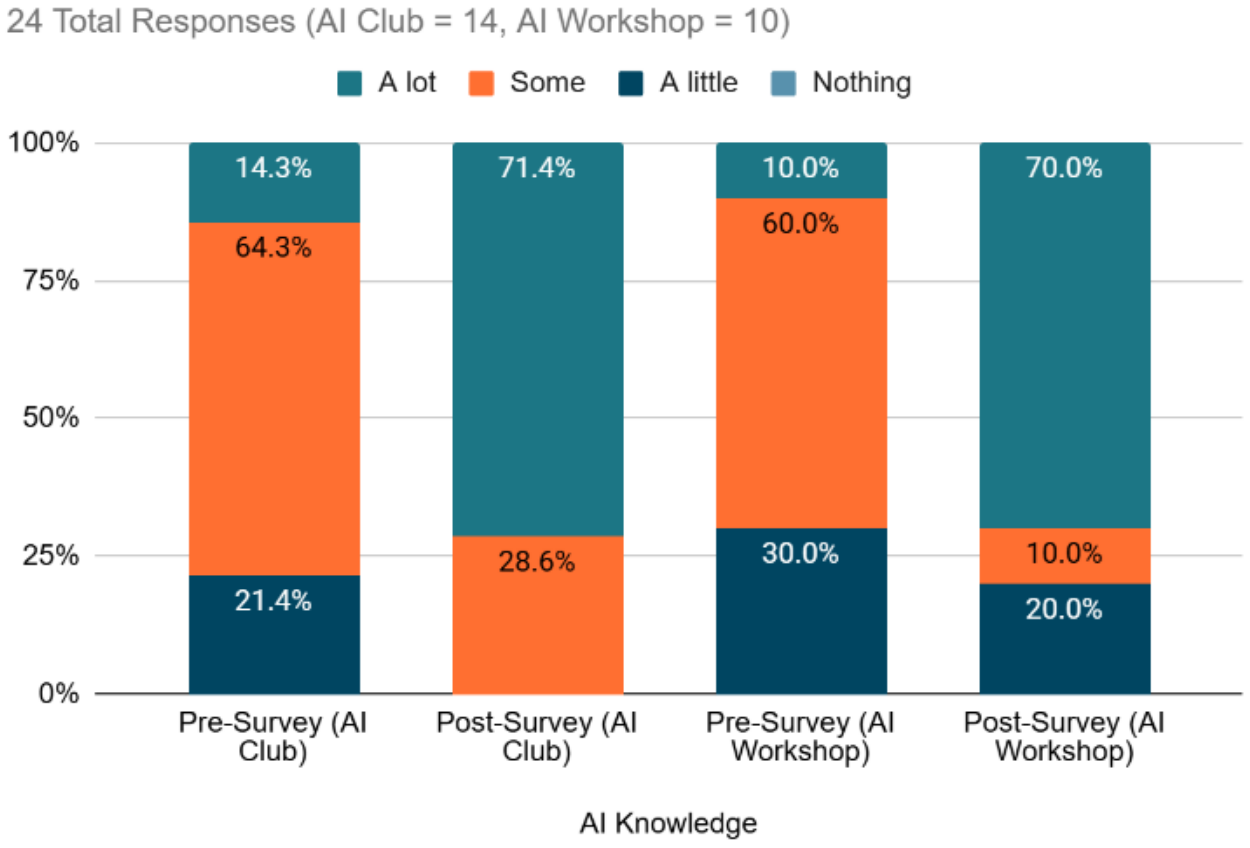

**Figure 2.** AI knowledge pre- and post-survey outcomes across two implementations.

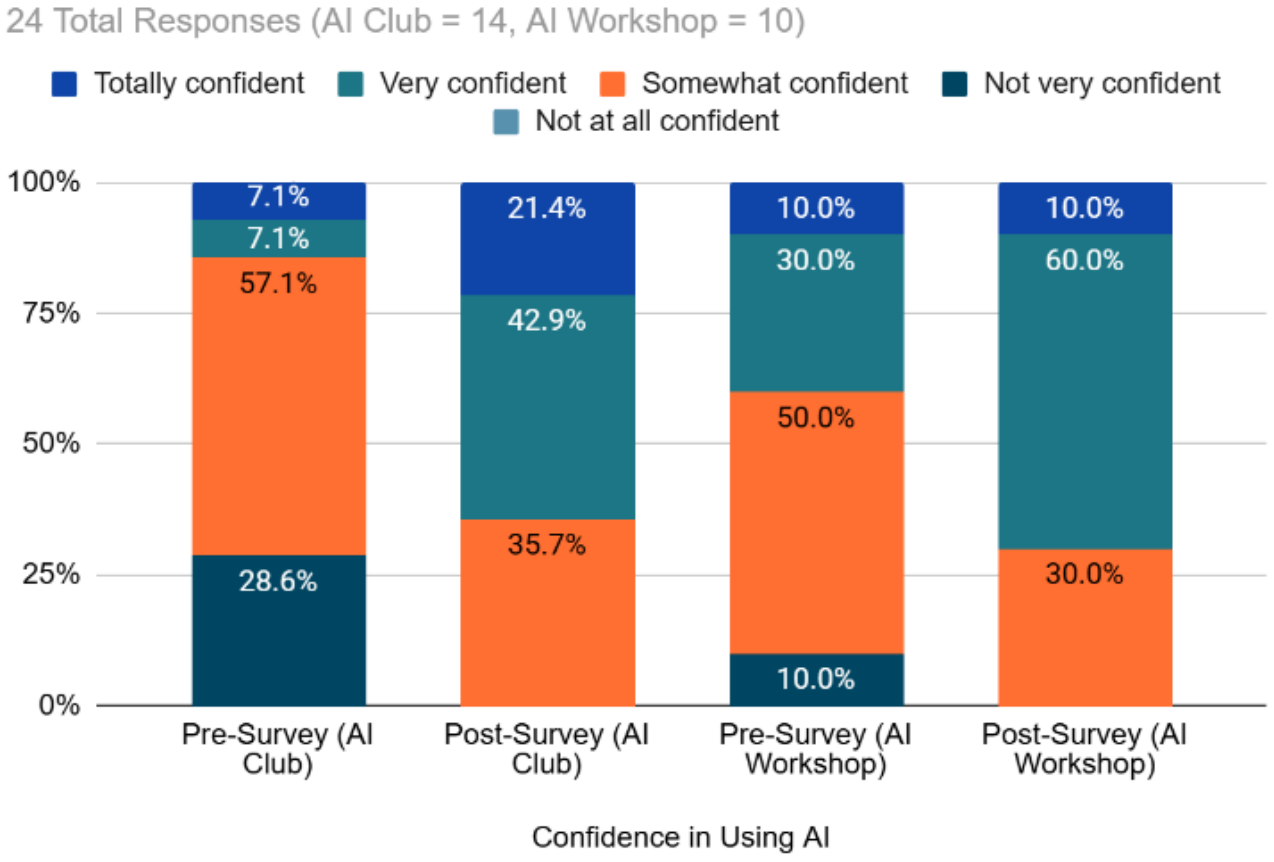

**Figure 3.** Confidence in using AI pre- and post-survey outcomes across two implementations.

In the virtual AI Innovation Club, self-reported AI knowledge shifted from moderate to high familiarity. The proportion of participants reporting "A lot" of AI knowledge increased from 14.3% before the program to 71.4% after the program, while lower knowledge categories decreased.

Confidence followed a similar pattern. Before the intervention, most participants reported being only "somewhat confident" or less in using AI tools. Afterward, responses shifted toward higher confidence levels, with more participants selecting "very confident" or "totally confident" and no participants selecting low-confidence categories.

The in-person workshop showed similar, though more moderate, shifts. Participants reporting high AI knowledge increased from 10.0% to 70.0%, and confidence responses also moved toward higher categories, with no participants reporting low confidence after the workshop. Together, these shifts suggest that hands-on interaction with AI tools helped improve perceived competence and technological confidence.

### 4.4. Indicators of Engagement and Inclusion

Participant reflections suggest that learners perceived the workshops as inclusive and supportive learning environments. Comments emphasizing inclusion, interactivity, collaborative pacing, and accessibility may indicate that culturally responsive facilitation contributed positively to participant engagement and comfort within STEM learning spaces. For example, one participant noted, "I liked how there was a great amount of inclusion, interactivity," while another stated, "Honestly, there were so many things I enjoyed, but one of them was how inclusive it was and how it was taken slowly for everyone to understand."

### 4.5. Conceptual & Ethical Understanding of AI

Post-survey results further indicate strong conceptual engagement and ethical awareness, particularly within the virtual AI Innovation Club (Figure 4). A large majority of participants agreed or strongly agreed that they could explain what artificial intelligence is and that they learned new things about how AI works, suggesting meaningful conceptual understanding beyond surface-level exposure.

Notably, ethical awareness outcomes were especially strong across participants. All respondents agreed or strongly agreed that AI can make mistakes and requires human oversight, indicating the development of critical AI literacy rather than purely functional knowledge. This finding is significant, as introductory AI workshops often emphasize tool usage while overlooking discussions of limitations, bias, and responsible AI practices. This emphasis on responsible AI was also reflected in participant comments, such as "Ethics and safety are important," indicating that students engaged not only with how AI works but also with its limitations and responsible use.

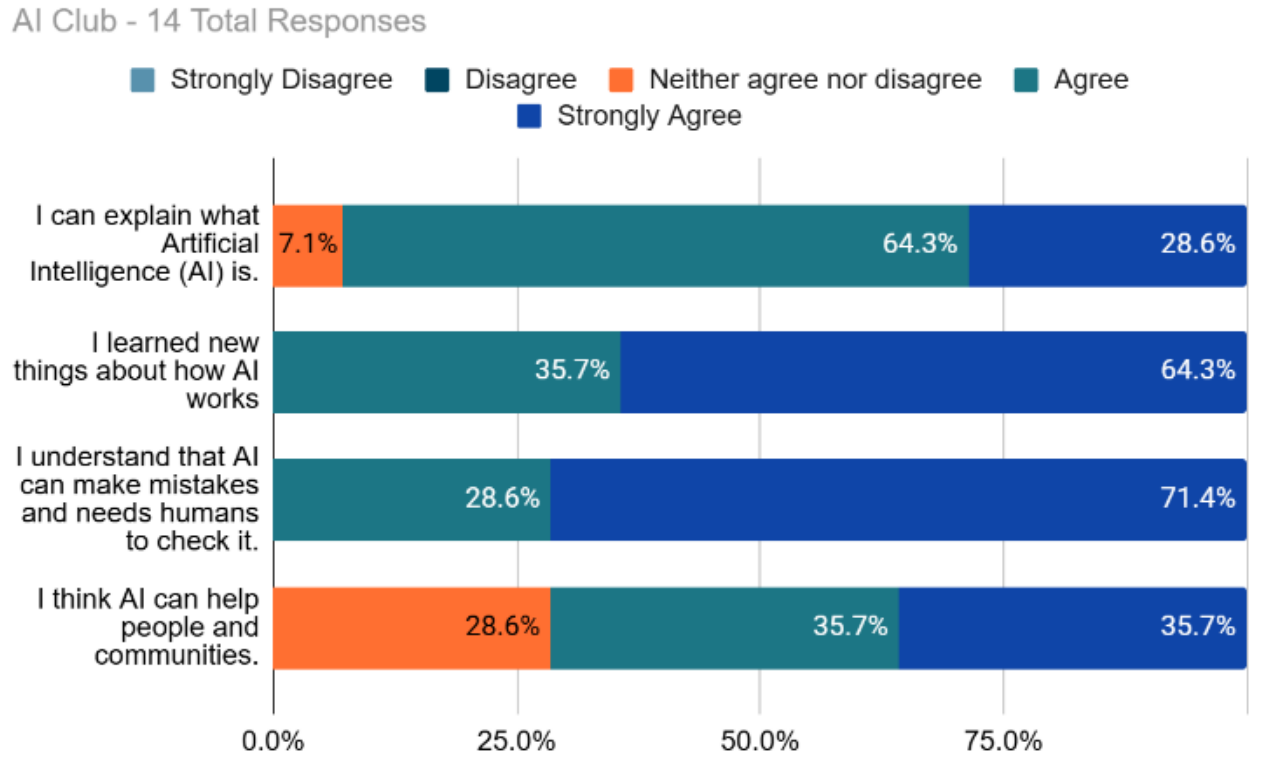

**Figure 4.** Post-survey agreement levels regarding conceptual and ethical understanding of AI in the virtual AI Innovation Club.

Participants also reported positive perceptions of AI's societal relevance, with most agreeing that AI can help people and communities. The inclusion of culturally affirming examples, real-world applications, and discussions of diverse innovators may have contributed to this broader understanding of AI's social impact.

### 4.6. Implications for a Culturally Responsive AI Education Framework

Taken together, the findings provide preliminary evidence that culturally responsive AI outreach may support short-term gains in AI literacy, confidence, and critical understanding within informal STEM learning environments. The consistent gains observed across both multi-session and short-duration implementations suggest that culturally relevant, hands-on AI learning experiences may support meaningful short-term outcomes even in time-constrained settings.

Notably, the clear improvements in confidence alongside knowledge gains indicate that culturally affirming and experiential learning environments may reduce technological intimidation and increase self-efficacy in emerging technologies. These findings are particularly relevant for youth with limited prior exposure to AI, as confidence has been associated with sustained engagement in STEM pathways [4]. Within informal outreach settings, confidence-building may serve as an important early indicator associated with continued participation and willingness to engage with emerging technologies.

Furthermore, the high levels of ethical awareness highlight a distinctive contribution of the framework: positioning youth not only as users of AI technologies but as critical thinkers who understand AI's limitations, societal implications, and responsible use. By integrating cultural relevance, mentorship, and interactive AI activities, the framework moves beyond traditional technical instruction toward a more holistic model of AI literacy that supports cognitive, affective, and critical learning outcomes simultaneously.

Overall, these results suggest that culturally responsive AI education may represent a promising and scalable approach for supporting measurable learning gains, confidence, cultural relevance, and critical awareness in inclusive youth STEM learning environments. Although the study did not directly measure long-term STEM belonging, identity development, or persistence intentions, the observed increases in confidence, engagement, and participant perceptions of inclusion suggest that culturally responsive AI outreach may support foundational conditions associated with continued participation in STEM learning.

### 4.7. Limitations

Several limitations should be considered when interpreting these findings. First, the study focused primarily on short-term AI literacy and confidence outcomes and did not directly measure constructs such as STEM belonging, identity development, long-term STEM persistence, or perceptions of systemic barriers. While participant reflections suggest positive experiences related to inclusion and engagement, these constructs were not formally assessed using validated belonging or identity measures. Second, the exploratory study design and relatively small sample size limit the generalizability of findings. Third, participant feedback was collected primarily through brief optional reflections rather than formal qualitative interviews or thematic analysis. As noted in related Canadian engineering outreach work, long-term outcomes such as career influence and sustained STEM pathway participation are difficult to measure within short-term outreach studies [16]. Therefore, the present findings should be interpreted as evidence of short-term AI literacy and confidence gains rather than definitive evidence of long-term belonging, empowerment, or STEM persistence. Future research should incorporate validated measures of belonging, STEM identity, empowerment, and persistence intentions alongside longitudinal follow-up to better understand how culturally responsive AI outreach may influence longer-term participation in STEM pathways.

## 5. CONCLUSIONS

Integrating AI literacy within culturally specific outreach showed both opportunities and challenges. Successes include increased engagement, greater enthusiasm for exploring technology, and participant reflections suggesting positive experiences within STEM learning environments. Challenges involve balancing technical complexity with accessibility, ensuring ongoing mentor training, and maintaining program sustainability. Overall, the findings suggest that culturally responsive and inclusive outreach programs may offer a promising approach for supporting early AI literacy, confidence, and engagement among Black youth in informal STEM learning environments. While additional research is needed to examine long-term impacts on STEM identity and participation, the study highlights the potential value of culturally affirming AI education initiatives in expanding access to emerging technological learning opportunities.

## Ethical Considerations

This study adhered to ethical research principles and involved secondary analysis of pre-existing outreach evaluation data collected for administrative and program improvement purposes. Survey responses were anonymous, and no personally identifiable information was collected or linked to participant responses. The study did not involve intervention, manipulation, or identification of participants, and findings are reported in aggregate form with anonymous illustrative comments.

## Disclosure Statement

ChatGPT (OpenAI GPT-5) was used in a limited capacity for language refinement and structural clarity. Grammarly was used for grammar review. AI tools were not used to generate research findings, conduct analysis, develop the study design, or create references. All conceptual development, data interpretation, and verification of sources were performed solely by the authors.